\documentclass[
tightenlines,
twocolumn,
a4paper,
nofootinbib,
superscriptaddress,
floatfix,
preprintnumbers,
flushbottom,
prl,
aps
]{revtex4-1}

\usepackage{url}
\usepackage[%
        hyperindex,breaklinks,
	pdfstartview=FitH,
  pdfpagemode=UseNone
	]{hyperref}

\usepackage[
   centertags, 
   sumlimits,  
   intlimits,  
   namelimits, 
]{amsmath} %
\usepackage{amssymb}

\makeatletter
\def\tagform@#1{\maketag@@@{\ignorespaces#1\unskip\@@italiccorr}}
\let\orgtheequation\theequation
\def\theequation{(\orgtheequation)}
\makeatother

\usepackage[all]{hypcap} 
\renewcommand{\BibitemShut}[1]{}

\usepackage[english]{babel} 

\usepackage[%
]{graphicx}

\def\l{\left}
\def\r{\right}

\begin{document}
\title{Elliptic flow and nuclear modification factor in ultra-relativistic heavy-ion collisions within a partonic transport model}

\author{Jan Uphoff}
 \email[E-mail: ]{uphoff@th.physik.uni-frankfurt.de}

\author{Oliver Fochler}

\author{Florian Senzel}

\author{Christian Wesp}
\affiliation{Institut f\"ur Theoretische Physik, Goethe-Universit\"at Frankfurt, Max-von-Laue-Str.\ 1, 
D-60438 Frankfurt am Main, Germany}

\author{Zhe Xu}
\affiliation{Department of Physics, Tsinghua University, Beijing 100084, China}

\author{Carsten Greiner}
\affiliation{Institut f\"ur Theoretische Physik, Goethe-Universit\"at Frankfurt, Max-von-Laue-Str.\ 1, 
D-60438 Frankfurt am Main, Germany}

\date{\today}

\begin{abstract}
The quark gluon plasma produced in ultra-relativistic heavy-ion collisions exhibits remarkable features. It behaves like a nearly perfect liquid with a small shear viscosity to entropy density ratio and leads to the quenching of highly energetic particles. We show that both effects can be understood for the first time within one common framework. Employing the parton cascade \emph{Boltzmann Approach to Multi-Parton Scatterings} (BAMPS), the microscopic interactions and the space-time evolution of the quark gluon plasma are calculated by solving the relativistic Boltzmann equation. Based on cross sections obtained from perturbative QCD  with explicitly taking the running coupling into account, we calculate the nuclear modification factor and elliptic flow in ultra-relativistic heavy-ion collisions. With only one single parameter associated with coherence effects of medium-induced gluon radiation, the experimental data of both observables can be understood on a microscopic level. Furthermore, we show that 
perturbative QCD interactions with a running coupling lead to a sufficiently small shear viscosity to entropy density ratio of the quark gluon plasma, which provides a microscopic explanation for the observations stated by hydrodynamic calculations.
\end{abstract}


\maketitle

\pagebreak

In ultra-relativistic heavy-ion collisions at the Relativistic Heavy-Ion Collider (RHIC) at BNL and the Large Hadron Collider (LHC) at CERN a hot and dense medium is created that consists of quarks and gluons. Experimental data shows that this quark gluon plasma (QGP) possesses a strong collective behavior and that high-energy partons deposit a sizeable amount of their energy in this medium \cite{Jacak:2012dx,Muller:2012zq}.

The collective behavior is often quantified by the elliptic flow coefficient $v_2$, which is the second harmonic of the Fourier decomposition of the azimuthal angle distribution of particle yields. Comparisons to hydrodynamic calculations reveal that the QGP behaves like a nearly perfect liquid with a small shear viscosity to entropy density ratio \cite{Gale:2013da}. However, the microscopic reason for this small ratio is currently not understood.

Experimental data of the nuclear modification factor $R_{AA}$, which is defined as the yield in heavy-ion (A+A) collisions divided by the yield in proton-proton (p+p) collisions scaled with the number of binary collisions,
\begin{align}
\label{raa_def}
  R_{AA}=\frac{{\rm d}^{2}N_{\text{AA}}/{\rm d}p_{T}{\rm d}y}{N_{\rm bin} \, {\rm d}^{2}N_{\text{pp}}/{\rm d}p_{T}{\rm d}y} \ ,
\end{align}
and the momentum imbalance of fully reconstructed jets indicate that high-energy particles are quenched by the created medium and lose lots of their energy \cite{Jacak:2012dx,Muller:2012zq}. Several calculations based on perturbative QCD (pQCD) energy loss in the QGP can describe the experimental data
\cite{Dainese:2004te,Vitev:2002pf,Vitev:2004bh,Salgado:2003gb,Armesto:2005iq,Wicks:2005gt,Schenke:2009gb,Renk:2011gj}.

A simultaneous understanding of collective bulk phenomena and jet quenching on the microscopic level remains a challenge, although several partonic transport models \cite{Geiger:1991nj, Zhang:1997ej,Molnar:2000jh,Bass:2002fh,Xu:2004mz,Lin:2004en} have been developed to address this issue. In this paper we will present new results on both observables obtained with the partonic transport model \emph{Boltzmann Approach to Multi-Parton Scatterings} (BAMPS). Based on cross sections calculated in pQCD, soft and hard particles are treated on the same footing in a common framework. While we take explicitly the running of the coupling into account,  we study not only the energy loss of highly energetic particles, but also the collective behavior of the bulk medium.

After a short introduction to BAMPS and the underlying physics, we address the employed pQCD cross sections and how the Landau-Pomeranchuk-Migdal (LPM) effect is implemented in our approach. Subsequently, we compare our results for $R_{AA}$ and $v_2$ with experimental data at RHIC and LHC and study the averaged value of the running coupling and the shear viscosity to entropy density ratio of the QGP for deeper insights in the properties of the hot and dense matter.


The partonic transport model \emph{Boltzmann Approach to Multi-Parton Scatterings} (BAMPS) \cite{Xu:2004mz,Xu:2007aa} describes the 3+1 dimensional evolution of the QGP phase by solving the Boltzmann equation for on-shell partons. All $2 \rightarrow 2$ \emph{and} $2 \leftrightarrow 3$ processes for light quarks (number of flavors $n_f=3$, $q=u,d,s$) and gluons ($g$) are included employing pQCD cross sections. In contrast to earlier BAMPS calculations, the coupling $\alpha_s$ is not assumed to be fixed, but its running is explicitly taken into account by setting the scale to the momentum transfer of the considered channel. This is done analogously to the implementation of heavy quarks in BAMPS \cite{Uphoff:2011ad,Uphoff:2012gb}.
The initial parton distributions are obtained from PYTHIA \cite{Sjostrand:2006za} and the Monte-Carlo Glauber model as described in detail in Ref.~\cite{Xu:2004mz,Uphoff:2010sh}. Initial event-by-event fluctuations are washed out by the employed testparticle method \cite{Xu:2004mz}.

In Ref.~\cite{Fochler:2013epa} we have recently presented an improved version of the Gunion-Bertsch (GB) matrix element for $2 \leftrightarrow 3$ processes, which cures problems of the original matrix element \cite{Gunion:1981qs} at forward and backward rapidity of the emitted gluon. Numerical comparisons to the exact matrix element show a good agreement \cite{Fochler:2013epa}.

Within the GB approximation the improved GB matrix element for the process $X\rightarrow Y + g$ factorizes in the binary matrix element for $X\rightarrow Y$ and a radiative factor $P_g$ \cite{Fochler:2013epa}
\begin{align}
\label{me_gb}
		{\l|\overline{\mathcal{M}}_{X\rightarrow Y + g}\r|}^2 =
	\l|\overline{\mathcal{M}}_{X\rightarrow Y}\r|^2 \, P_g
\end{align}
with
\begin{multline}
\label{gb_pgm_radiation_spectrum}
	P_g = 48 \pi \alpha_s(k_\perp^2) \, (1-\bar{x})^2  \,	\\
  \times\, \l[ \frac{ {\bf k}_\perp}{k_\perp^2} +  \frac{ {\bf q}_\perp - {\bf k}_\perp }{ ({\bf q}_\perp - {\bf k}_\perp)^2+ m_D^2\left(\alpha_s(k_\perp^2)\right)} \r]^2 \ .
\end{multline}
The transverse momentum of the emitted  and internal gluons are denoted with ${\bf k}_\perp$ and ${\bf q}_\perp$, respectively. The longitudinal momentum fraction $\bar x$ is related to the rapidity of the emitted gluon via $\bar x = k_\perp e^{|y|}/\sqrt{s}$, where $s$ is the squared center of mass energy of the interaction.
$X\rightarrow Y$ stand for any binary process of light quarks and gluons, while only (Mandelstam) $t$ channel dominated processes (equivalent to $X=Y$) have a finite contribution within the GB approximation. These binary matrix elements are given in the same approximation by
\begin{align}
\label{me_22}
		{\l|\overline{\mathcal{M}}_{X\rightarrow Y}\r|}^2 =
	C_{X\rightarrow Y} \,64  \pi^2 \alpha_s^2(t)	 \frac{s^2}{[t-m_{D}^2(\alpha_s(t))]^2} \ ,
\end{align}
where $C_{X\rightarrow Y}$ is the color factor of the respective process. 
All internal gluon propagators in Eqs.~\ref{gb_pgm_radiation_spectrum} and \ref{me_22} are screened with the Debye mass $m_D$.
The coupling in the definition of the Debye mass is also evaluated at the respective scale of the propagators.
The first term in the bracket in Eq.~\ref{gb_pgm_radiation_spectrum} does not need to be screened by a screening mass since it corresponds to the external emitted gluon and the infrared divergence is cured by the implementation of the LPM effect in BAMPS \cite{Xu:2004mz}. Since including coherence effects consistently in a semi-classical transport model is difficult, the LPM suppression is effectively implemented by only allowing completely independent scatterings, demanding that the formation time~$\tau$ of the emitted gluon is smaller than the mean free path~$\lambda$ of the emitting particle. To this end, the LPM effect is included in BAMPS via a $\Theta$~function in the integrand of the $2\leftrightarrow 3$ cross section, \cite{Xu:2004mz}
\begin{align}
  \Theta\left( \lambda - X \tau \right) \ .
\end{align}
$X=0$ corresponds to no LPM suppression while $X=1$, the previously \cite{Xu:2004mz} used value in BAMPS, discards all interfering processes altogether. We expect that a more sophisticated treatment of the LPM effect would allow also some interference processes, leading effectively to $0 <X<1$. Thus, we treat $X$ as a parameter and calibrate its value to the nuclear modification factor of neutral pions at RHIC.

In the following we present an update of previous calculations for the elliptic flow \cite{Xu:2007jv,Xu:2008av,Xu:2010cq,Fochler:2011en} and the nuclear modification factor \cite{Fochler:2008ts,Fochler:2010wn,Fochler:2011en}, now including the improved GB cross section as well as a running coupling for all channels.

\begin{figure}
\includegraphics[width=1.0\linewidth]{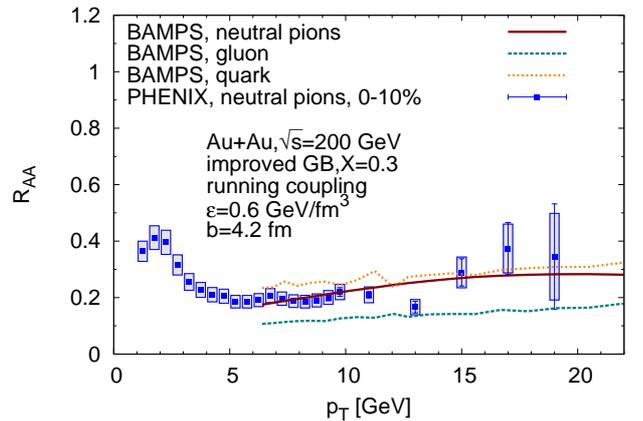}
\caption{Nuclear modification factor $R_{AA}$ of gluons, light quarks, and neutral pions at RHIC for a running coupling and LPM parameter $X=0.3$ together with data of neutral pions \cite{Adare:2008qa}.}
\label{fig:raa_rhic}
\end{figure}

Figure~\ref{fig:raa_rhic} depicts the nuclear modification factor of light partons and neutral pions at RHIC.
Due to the larger color factor, gluons are considerably stronger suppressed than light quarks.
For comparing with experimental data of neutral pions, we perform the fragmentation of gluons and light quarks to neutral pions with the AKK fragmentation functions \cite{Albino:2008fy}. The pion curve lies between the gluon and light quark curve. At small $p_T$ the pions are dominated by fragmentation from gluons, at large $p_T$ from light quarks. The LPM parameter $X$ is chosen as $X=0.3$ to give the best agreement with the data. In the following we keep this parameter fixed and compare to other experimental data of the $R_{AA}$ and $v_2$ at RHIC and LHC. As we explained above, an $X$ smaller than one should be consistent with a more sophisticated LPM treatment in BAMPS. 

\begin{figure}
\includegraphics[ width=1.0\linewidth]{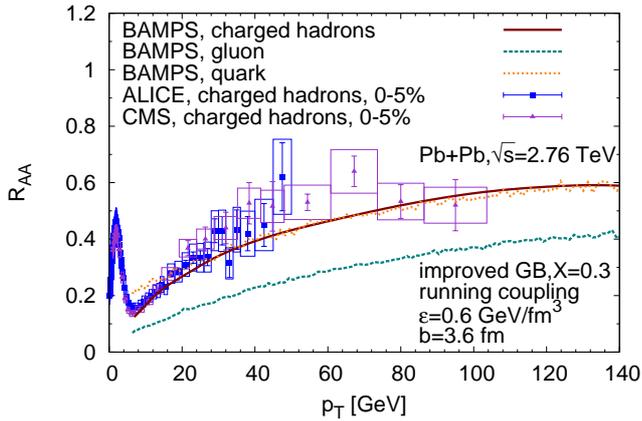}
\caption{Nuclear modification factor $R_{AA}$ of gluons, light quarks, and charged hadrons at LHC for a running coupling  and LPM parameter $X=0.3$ together with data of charged hadrons \cite{Abelev:2012hxa}.}
\label{fig:raa_lhc}
\end{figure}

With $X=0.3$ we find a very good agreement with the experimental data at RHIC. The same holds at LHC, as is shown in Fig.~\ref{fig:raa_lhc}.
In addition to the gluon and light quark curves, we also depict the curve for charged hadrons obtained again via AKK fragmentation. Again, the hadron curve lies mainly between the light quark and gluon curves, but is slightly larger than the light quark curve for large~$p_T$. Due to the fragmentation process a hadron possesses on average a transverse momentum of only about half of its parental light parton, that is, the charged hadron $R_{AA}$ at a given $p_T$ has approximately the same value as the parton $R_{AA}$ at twice as large $p_T$. Due to the rise in the $R_{AA}$ the hadron curve is shifted to larger values than the parton curve at same $p_T$.
Furthermore, the slope of the hadron curve at intermediate $p_T$ is steeper than the parton curves due to the fragmentation process as well as the different slopes of the gluon and light quark spectra. Hadrons at large $p_T$ are dominated by quark fragmentation, while hadrons at small $p_T$ are dominated by gluon fragmentation. 

Having presented the results for high-energy particles, we now address the bulk medium interactions. It is important to note that all partons in BAMPS are treated on the same footing, that is, all particles interact based on the pQCD cross sections introduced above. Since hadronization from the partonic to the hadronic phase is not well understood in the soft regime, we compare the \emph{integrated} $v_2$ on the parton level to experimental data, as the integrated $v_2$ should not be sensitive to the phase transition. 

\begin{figure}
\includegraphics[angle=270, width=1.0\linewidth]{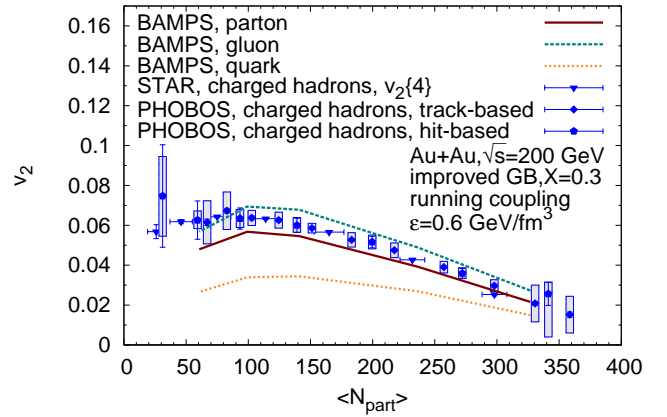}
\caption{Elliptic flow $v_2$ of gluons, light quarks, and both together (light partons) within $|\eta|<1.0$ as a function of the number of participants $N_{\rm part}$ at RHIC for a running coupling and LPM parameter $X=0.3$. As a comparison we show experimental data by STAR and PHOBOS for charged hadrons within $|\eta|<0.5$ and $|\eta|<1.0$ \cite{Adams:2004bi,Back:2004mh}.}
\label{fig:v2_rhic}
\end{figure}

\begin{figure}
\includegraphics[angle=270,  width=1.0\linewidth]{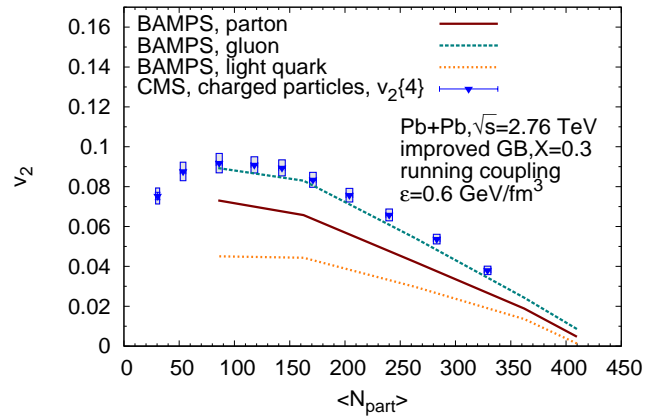}
\caption{Elliptic flow $v_2$ of gluons, light quarks, and both together (light partons) within $|\eta|<0.8$ as a function of the number of participants $N_{\rm part}$ at LHC for a running coupling and LPM parameter $X=0.3$. As a comparison we show experimental data by CMS for charged hadrons within $|\eta|<0.8$ \cite{Chatrchyan:2012ta}.}
\label{fig:v2_lhc}
\end{figure}

In Fig.~\ref{fig:v2_rhic} and \ref{fig:v2_lhc} the integrated elliptic flow is shown as a function of the number of participants at RHIC and LHC, respectively.
With the same parameter $X=0.3$ we obtain a sizeable elliptic flow on the parton level, which is calculated after a partonic freeze-out energy density of $\epsilon = 0.6 \, {\rm GeV/fm^3}$ is reached \cite{Xu:2008av}. 
The gluon elliptic flow is close to the data and the light quark $v_2$ smaller due to the smaller color factor. The integrated $v_2$ of all light partons is the curve that should be compared to the data. It is slightly smaller than the data since hadronic final interactions and event-by-event fluctuations in the initial state are not taken into account. As shown in Ref.~\cite{Auvinen:2013sba} hadronic contributions may increase the $v_2$ by about $10-15$\,\%, which could explain part of the small deviation between the light parton curve and the experimental data.
Furthermore, explicit consideration of quantum statistics could increase the elliptic flow due to Bose enhancement of gluons while the nuclear modification factor of high energy particles is not influenced.

\begin{figure}
\includegraphics[width=1.0\linewidth]{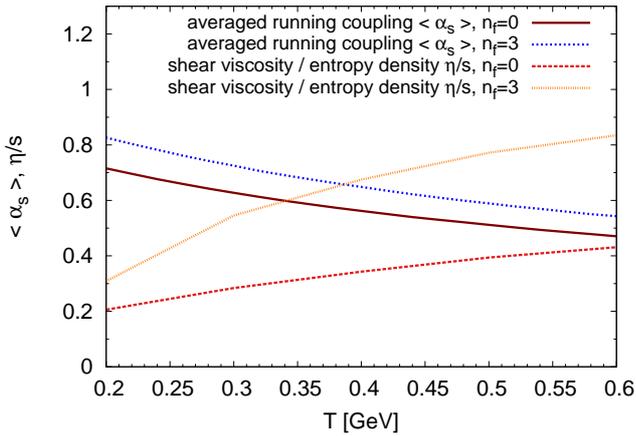}
\caption{Shear viscosity over entropy density~$\eta/s$ for running coupling and $X=0.3$ in a static medium of temperature~$T$ with number of quark flavors~$n_f$. Furthermore, the average value of the running coupling for binary collisions in a thermal medium of temperature~$T$ is shown.}
\label{fig:eta_s_alpha}
\end{figure}

It is a remarkable result that we obtain a sizeable $v_2$ while having the same suppression as the experimental data at large $p_T$. The reason for this lies partly in the isotropization of inelastic $2 \leftrightarrow 3$ processes and partly in the running coupling. For particles with small $p_T$ the coupling is on average stronger as for high-energy particles, which affects the elliptic flow at small $p_T$ and $R_{AA}$ at large $p_T$ differently.
In Fig.~\ref{fig:eta_s_alpha} the averaged running coupling of binary collisional processes is depicted in a static thermal medium as a function of the temperature of the medium, where the coupling is evaluated microscopically at the momentum transfer of each interaction.
As expected, the average coupling decreases with increasing temperature.

As advocated in dissipative hydrodynamic fits, an important quantity for the bulk medium in heavy-ion collisions is the shear viscosity to entropy density ratio~$\eta/s$. In Fig.~\ref{fig:eta_s_alpha} the temperature dependence of this value in a static medium allowing all $2\rightarrow 2$ and $2\leftrightarrow 3$ is shown. 
The shear viscosity is calculated via the Green-Kubo relation, which links
the autocorrelation function of the medium energy-momentum tensor of the medium to the
transport coefficient $\eta$ \cite{Wesp:2011yy}.
The ratio $\eta/s$ decreases with decreasing temperature and reaches a minimum at the phase transition. The region that is most relevant for the elliptic flow lies around $T=0.2 \, {\rm GeV}$ for $n_f=0$ (the QGP at RHIC and LHC is mostly gluon dominated; in our calculation the gluon and quark fugacity at freeze-out at LHC take values of approximately 0.9 and 0.5, respectively). Here, the value of $\eta/s$ is approximately $0.2$, which agrees very well with the shear viscosity extraction from dissipative hydrodynamic models \cite{Gale:2012rq}. Thus our calculation employing pQCD cross sections can give a microscopic explanation of the small shear viscosity to entropy density ratio extracted from hydrodynamics.

In summary, we compared results on the nuclear modification factor and elliptic flow in ultra-relativistic heavy-ion collisions obtained from full microscopic, non-equilibrium transport calculations to experimental data. With pQCD cross sections and a running coupling the experimental data can be understood on the parton level, although some contribution to the elliptic flow from hadronic interactions might be relevant. We show that these interactions lead to a sufficiently small shear viscosity to entropy density ratio and thus can provide a microscopic explanation of the small value extracted from viscous hydrodynamics. As a future project, it would be interesting to study possible improvements on the LPM effect \cite{Zapp:2008af, ColemanSmith:2012vr} and include quantum statistics instead of Boltzmann statistics. Moreover, we will further investigate the energy loss of fully reconstructed jets \cite{Senzel:2013dta} and heavy flavor \cite{Uphoff:2013rka,Abir:2011jb} within the same framework.

 \textbf{Acknowledgments}: 
This work was supported by the Bundesministerium f\"ur Bildung und Forschung (BMBF), the NSFC under grant No.\ 11275103, HGS-HIRe and the Helmholtz International Center for FAIR within the framework of the LOEWE program launched by the State of Hesse. Numerical computations have been performed at the Center for Scientific Computing (CSC).

\bibliography{hq,text}

\begin{thebibliography}{44}%
\makeatletter
\providecommand \@ifxundefined [1]{%
 \@ifx{#1\undefined}
}%
\providecommand \@ifnum [1]{%
 \ifnum #1\expandafter \@firstoftwo
 \else \expandafter \@secondoftwo
 \fi
}%
\providecommand \@ifx [1]{%
 \ifx #1\expandafter \@firstoftwo
 \else \expandafter \@secondoftwo
 \fi
}%
\providecommand \natexlab [1]{#1}%
\providecommand \enquote  [1]{``#1''}%
\providecommand \bibnamefont  [1]{#1}%
\providecommand \bibfnamefont [1]{#1}%
\providecommand \citenamefont [1]{#1}%
\providecommand \href@noop [0]{\@secondoftwo}%
\providecommand \href [0]{\begingroup \@sanitize@url \@href}%
\providecommand \@href[1]{\@@startlink{#1}\@@href}%
\providecommand \@@href[1]{\endgroup#1\@@endlink}%
\providecommand \@sanitize@url [0]{\catcode `\\12\catcode `\$12\catcode
  `\&12\catcode `\#12\catcode `\^12\catcode `\_12\catcode `\%12\relax}%
\providecommand \@@startlink[1]{}%
\providecommand \@@endlink[0]{}%
\providecommand \url  [0]{\begingroup\@sanitize@url \@url }%
\providecommand \@url [1]{\endgroup\@href {#1}{\urlprefix }}%
\providecommand \urlprefix  [0]{URL }%
\providecommand \Eprint [0]{\href }%
\providecommand \doibase [0]{http://dx.doi.org/}%
\providecommand \selectlanguage [0]{\@gobble}%
\providecommand \bibinfo  [0]{\@secondoftwo}%
\providecommand \bibfield  [0]{\@secondoftwo}%
\providecommand \translation [1]{[#1]}%
\providecommand \BibitemOpen [0]{}%
\providecommand \bibitemStop [0]{}%
\providecommand \bibitemNoStop [0]{.\EOS\space}%
\providecommand \EOS [0]{\spacefactor3000\relax}%
\providecommand \BibitemShut  [1]{\csname bibitem#1\endcsname}%
\let\auto@bib@innerbib\@empty
\bibitem [{\citenamefont {Jacak}\ and\ \citenamefont
  {M\"uller}(2012)}]{Jacak:2012dx}%
  \BibitemOpen
  \bibfield  {author} {\bibinfo {author} {\bibfnamefont {B.~V.}\ \bibnamefont
  {Jacak}}\ and\ \bibinfo {author} {\bibfnamefont {B.}~\bibnamefont
  {M\"uller}},\ }\href {\doibase 10.1126/science.1215901} {\bibfield  {journal}
  {\bibinfo  {journal} {Science}\ }\textbf {\bibinfo {volume} {337}},\ \bibinfo
  {pages} {310} (\bibinfo {year} {2012})}\BibitemShut {NoStop}%
\bibitem [{\citenamefont {M\"uller}\ \emph {et~al.}(2012)\citenamefont
  {M\"uller}, \citenamefont {Schukraft},\ and\ \citenamefont
  {Wyslouch}}]{Muller:2012zq}%
  \BibitemOpen
  \bibfield  {author} {\bibinfo {author} {\bibfnamefont {B.}~\bibnamefont
  {M\"uller}}, \bibinfo {author} {\bibfnamefont {J.}~\bibnamefont {Schukraft}},
  \ and\ \bibinfo {author} {\bibfnamefont {B.}~\bibnamefont {Wyslouch}},\
  }\href {\doibase 10.1146/annurev-nucl-102711-094910} {\bibfield  {journal}
  {\bibinfo  {journal} {Ann.Rev.Nucl.Part.Sci.}\ }\textbf {\bibinfo {volume}
  {62}},\ \bibinfo {pages} {361} (\bibinfo {year} {2012})},\ \Eprint
  {http://arxiv.org/abs/1202.3233} {arXiv:1202.3233 [hep-ex]} \BibitemShut
  {NoStop}%
\bibitem [{\citenamefont {Gale}\ \emph
  {et~al.}(2013{\natexlab{a}})\citenamefont {Gale}, \citenamefont {Jeon},\ and\
  \citenamefont {Schenke}}]{Gale:2013da}%
  \BibitemOpen
  \bibfield  {author} {\bibinfo {author} {\bibfnamefont {C.}~\bibnamefont
  {Gale}}, \bibinfo {author} {\bibfnamefont {S.}~\bibnamefont {Jeon}}, \ and\
  \bibinfo {author} {\bibfnamefont {B.}~\bibnamefont {Schenke}},\ }\href
  {\doibase 10.1142/S0217751X13400113} {\bibfield  {journal} {\bibinfo
  {journal} {Int.J.Mod.Phys.}\ }\textbf {\bibinfo {volume} {A28}},\ \bibinfo
  {pages} {1340011} (\bibinfo {year} {2013}{\natexlab{a}})},\ \Eprint
  {http://arxiv.org/abs/1301.5893} {arXiv:1301.5893 [nucl-th]} \BibitemShut
  {NoStop}%
\bibitem [{\citenamefont {Dainese}\ \emph {et~al.}(2005)\citenamefont
  {Dainese}, \citenamefont {Loizides},\ and\ \citenamefont
  {Paic}}]{Dainese:2004te}%
  \BibitemOpen
  \bibfield  {author} {\bibinfo {author} {\bibfnamefont {A.}~\bibnamefont
  {Dainese}}, \bibinfo {author} {\bibfnamefont {C.}~\bibnamefont {Loizides}}, \
  and\ \bibinfo {author} {\bibfnamefont {G.}~\bibnamefont {Paic}},\ }\href
  {\doibase 10.1140/epjc/s2004-02077-x} {\bibfield  {journal} {\bibinfo
  {journal} {Eur.Phys.J.}\ }\textbf {\bibinfo {volume} {C38}},\ \bibinfo
  {pages} {461} (\bibinfo {year} {2005})},\ \Eprint
  {http://arxiv.org/abs/hep-ph/0406201} {arXiv:hep-ph/0406201} \BibitemShut
  {NoStop}%
\bibitem [{\citenamefont {Vitev}\ and\ \citenamefont
  {Gyulassy}(2002)}]{Vitev:2002pf}%
  \BibitemOpen
  \bibfield  {author} {\bibinfo {author} {\bibfnamefont {I.}~\bibnamefont
  {Vitev}}\ and\ \bibinfo {author} {\bibfnamefont {M.}~\bibnamefont
  {Gyulassy}},\ }\href {\doibase 10.1103/PhysRevLett.89.252301} {\bibfield
  {journal} {\bibinfo  {journal} {Phys.Rev.Lett.}\ }\textbf {\bibinfo {volume}
  {89}},\ \bibinfo {pages} {252301} (\bibinfo {year} {2002})},\ \Eprint
  {http://arxiv.org/abs/hep-ph/0209161} {arXiv:hep-ph/0209161 [hep-ph]}
  \BibitemShut {NoStop}%
\bibitem [{\citenamefont {Vitev}(2004)}]{Vitev:2004bh}%
  \BibitemOpen
  \bibfield  {author} {\bibinfo {author} {\bibfnamefont {I.}~\bibnamefont
  {Vitev}},\ }\href {\doibase 10.1088/0954-3899/30/8/019} {\bibfield  {journal}
  {\bibinfo  {journal} {J.Phys.}\ }\textbf {\bibinfo {volume} {G30}},\ \bibinfo
  {pages} {S791} (\bibinfo {year} {2004})},\ \Eprint
  {http://arxiv.org/abs/hep-ph/0403089} {arXiv:hep-ph/0403089 [hep-ph]}
  \BibitemShut {NoStop}%
\bibitem [{\citenamefont {Salgado}\ and\ \citenamefont
  {Wiedemann}(2003)}]{Salgado:2003gb}%
  \BibitemOpen
  \bibfield  {author} {\bibinfo {author} {\bibfnamefont {C.~A.}\ \bibnamefont
  {Salgado}}\ and\ \bibinfo {author} {\bibfnamefont {U.~A.}\ \bibnamefont
  {Wiedemann}},\ }\href {\doibase 10.1103/PhysRevD.68.014008} {\bibfield
  {journal} {\bibinfo  {journal} {Phys.Rev.}\ }\textbf {\bibinfo {volume}
  {D68}},\ \bibinfo {pages} {014008} (\bibinfo {year} {2003})},\ \Eprint
  {http://arxiv.org/abs/hep-ph/0302184} {arXiv:hep-ph/0302184 [hep-ph]}
  \BibitemShut {NoStop}%
\bibitem [{\citenamefont {Armesto}\ \emph {et~al.}(2005)\citenamefont
  {Armesto}, \citenamefont {Dainese}, \citenamefont {Salgado},\ and\
  \citenamefont {Wiedemann}}]{Armesto:2005iq}%
  \BibitemOpen
  \bibfield  {author} {\bibinfo {author} {\bibfnamefont {N.}~\bibnamefont
  {Armesto}}, \bibinfo {author} {\bibfnamefont {A.}~\bibnamefont {Dainese}},
  \bibinfo {author} {\bibfnamefont {C.~A.}\ \bibnamefont {Salgado}}, \ and\
  \bibinfo {author} {\bibfnamefont {U.~A.}\ \bibnamefont {Wiedemann}},\ }\href
  {\doibase 10.1103/PhysRevD.71.054027} {\bibfield  {journal} {\bibinfo
  {journal} {Phys.Rev.}\ }\textbf {\bibinfo {volume} {D71}},\ \bibinfo {pages}
  {054027} (\bibinfo {year} {2005})},\ \Eprint
  {http://arxiv.org/abs/hep-ph/0501225} {arXiv:hep-ph/0501225 [hep-ph]}
  \BibitemShut {NoStop}%
\bibitem [{\citenamefont {Wicks}\ \emph {et~al.}(2007)\citenamefont {Wicks},
  \citenamefont {Horowitz}, \citenamefont {Djordjevic},\ and\ \citenamefont
  {Gyulassy}}]{Wicks:2005gt}%
  \BibitemOpen
  \bibfield  {author} {\bibinfo {author} {\bibfnamefont {S.}~\bibnamefont
  {Wicks}}, \bibinfo {author} {\bibfnamefont {W.}~\bibnamefont {Horowitz}},
  \bibinfo {author} {\bibfnamefont {M.}~\bibnamefont {Djordjevic}}, \ and\
  \bibinfo {author} {\bibfnamefont {M.}~\bibnamefont {Gyulassy}},\ }\href
  {\doibase 10.1016/j.nuclphysa.2006.12.048} {\bibfield  {journal} {\bibinfo
  {journal} {Nucl.Phys.}\ }\textbf {\bibinfo {volume} {A784}},\ \bibinfo
  {pages} {426} (\bibinfo {year} {2007})},\ \Eprint
  {http://arxiv.org/abs/nucl-th/0512076} {arXiv:nucl-th/0512076} \BibitemShut
  {NoStop}%
\bibitem [{\citenamefont {Schenke}\ \emph {et~al.}(2009)\citenamefont
  {Schenke}, \citenamefont {Gale},\ and\ \citenamefont
  {Jeon}}]{Schenke:2009gb}%
  \BibitemOpen
  \bibfield  {author} {\bibinfo {author} {\bibfnamefont {B.}~\bibnamefont
  {Schenke}}, \bibinfo {author} {\bibfnamefont {C.}~\bibnamefont {Gale}}, \
  and\ \bibinfo {author} {\bibfnamefont {S.}~\bibnamefont {Jeon}},\ }\href
  {\doibase 10.1103/PhysRevC.80.054913} {\bibfield  {journal} {\bibinfo
  {journal} {Phys.Rev.}\ }\textbf {\bibinfo {volume} {C80}},\ \bibinfo {pages}
  {054913} (\bibinfo {year} {2009})},\ \Eprint {http://arxiv.org/abs/0909.2037}
  {arXiv:0909.2037 [hep-ph]} \BibitemShut {NoStop}%
\bibitem [{\citenamefont {Renk}\ \emph {et~al.}(2011)\citenamefont {Renk},
  \citenamefont {Holopainen}, \citenamefont {Paatelainen},\ and\ \citenamefont
  {Eskola}}]{Renk:2011gj}%
  \BibitemOpen
  \bibfield  {author} {\bibinfo {author} {\bibfnamefont {T.}~\bibnamefont
  {Renk}}, \bibinfo {author} {\bibfnamefont {H.}~\bibnamefont {Holopainen}},
  \bibinfo {author} {\bibfnamefont {R.}~\bibnamefont {Paatelainen}}, \ and\
  \bibinfo {author} {\bibfnamefont {K.~J.}\ \bibnamefont {Eskola}},\ }\href
  {\doibase 10.1103/PhysRevC.84.014906} {\bibfield  {journal} {\bibinfo
  {journal} {Phys.Rev.}\ }\textbf {\bibinfo {volume} {C84}},\ \bibinfo {pages}
  {014906} (\bibinfo {year} {2011})},\ \Eprint {http://arxiv.org/abs/1103.5308}
  {arXiv:1103.5308 [hep-ph]} \BibitemShut {NoStop}%
\bibitem [{\citenamefont {Geiger}\ and\ \citenamefont
  {M\"uller}(1992)}]{Geiger:1991nj}%
  \BibitemOpen
  \bibfield  {author} {\bibinfo {author} {\bibfnamefont {K.}~\bibnamefont
  {Geiger}}\ and\ \bibinfo {author} {\bibfnamefont {B.}~\bibnamefont
  {M\"uller}},\ }\href {\doibase 10.1016/0550-3213(92)90280-O} {\bibfield
  {journal} {\bibinfo  {journal} {Nucl.Phys.}\ }\textbf {\bibinfo {volume}
  {B369}},\ \bibinfo {pages} {600} (\bibinfo {year} {1992})}\BibitemShut
  {NoStop}%
\bibitem [{\citenamefont {Zhang}(1998)}]{Zhang:1997ej}%
  \BibitemOpen
  \bibfield  {author} {\bibinfo {author} {\bibfnamefont {B.}~\bibnamefont
  {Zhang}},\ }\href {\doibase 10.1016/S0010-4655(98)00010-1} {\bibfield
  {journal} {\bibinfo  {journal} {Comput.Phys.Commun.}\ }\textbf {\bibinfo
  {volume} {109}},\ \bibinfo {pages} {193} (\bibinfo {year} {1998})},\ \Eprint
  {http://arxiv.org/abs/nucl-th/9709009} {arXiv:nucl-th/9709009} \BibitemShut
  {NoStop}%
\bibitem [{\citenamefont {Molnar}\ and\ \citenamefont
  {Gyulassy}(2000)}]{Molnar:2000jh}%
  \BibitemOpen
  \bibfield  {author} {\bibinfo {author} {\bibfnamefont {D.}~\bibnamefont
  {Molnar}}\ and\ \bibinfo {author} {\bibfnamefont {M.}~\bibnamefont
  {Gyulassy}},\ }\href {\doibase 10.1103/PhysRevC.62.054907} {\bibfield
  {journal} {\bibinfo  {journal} {Phys.Rev.}\ }\textbf {\bibinfo {volume}
  {C62}},\ \bibinfo {pages} {054907} (\bibinfo {year} {2000})},\ \Eprint
  {http://arxiv.org/abs/nucl-th/0005051} {arXiv:nucl-th/0005051 [nucl-th]}
  \BibitemShut {NoStop}%
\bibitem [{\citenamefont {Bass}\ \emph {et~al.}(2003)\citenamefont {Bass},
  \citenamefont {M\"uller},\ and\ \citenamefont {Srivastava}}]{Bass:2002fh}%
  \BibitemOpen
  \bibfield  {author} {\bibinfo {author} {\bibfnamefont {S.}~\bibnamefont
  {Bass}}, \bibinfo {author} {\bibfnamefont {B.}~\bibnamefont {M\"uller}}, \
  and\ \bibinfo {author} {\bibfnamefont {D.}~\bibnamefont {Srivastava}},\
  }\href {\doibase 10.1016/S0370-2693(02)03068-X} {\bibfield  {journal}
  {\bibinfo  {journal} {Phys.Lett.}\ }\textbf {\bibinfo {volume} {B551}},\
  \bibinfo {pages} {277} (\bibinfo {year} {2003})},\ \Eprint
  {http://arxiv.org/abs/nucl-th/0207042} {arXiv:nucl-th/0207042 [nucl-th]}
  \BibitemShut {NoStop}%
\bibitem [{\citenamefont {Xu}\ and\ \citenamefont {Greiner}(2005)}]{Xu:2004mz}%
  \BibitemOpen
  \bibfield  {author} {\bibinfo {author} {\bibfnamefont {Z.}~\bibnamefont
  {Xu}}\ and\ \bibinfo {author} {\bibfnamefont {C.}~\bibnamefont {Greiner}},\
  }\href {\doibase 10.1103/PhysRevC.71.064901} {\bibfield  {journal} {\bibinfo
  {journal} {Phys.Rev.}\ }\textbf {\bibinfo {volume} {C71}},\ \bibinfo {pages}
  {064901} (\bibinfo {year} {2005})},\ \Eprint
  {http://arxiv.org/abs/hep-ph/0406278} {arXiv:hep-ph/0406278} \BibitemShut
  {NoStop}%
\bibitem [{\citenamefont {Lin}\ \emph {et~al.}(2005)\citenamefont {Lin},
  \citenamefont {Ko}, \citenamefont {Li}, \citenamefont {Zhang},\ and\
  \citenamefont {Pal}}]{Lin:2004en}%
  \BibitemOpen
  \bibfield  {author} {\bibinfo {author} {\bibfnamefont {Z.-W.}\ \bibnamefont
  {Lin}}, \bibinfo {author} {\bibfnamefont {C.~M.}\ \bibnamefont {Ko}},
  \bibinfo {author} {\bibfnamefont {B.-A.}\ \bibnamefont {Li}}, \bibinfo
  {author} {\bibfnamefont {B.}~\bibnamefont {Zhang}}, \ and\ \bibinfo {author}
  {\bibfnamefont {S.}~\bibnamefont {Pal}},\ }\href {\doibase
  10.1103/PhysRevC.72.064901} {\bibfield  {journal} {\bibinfo  {journal}
  {Phys.Rev.}\ }\textbf {\bibinfo {volume} {C72}},\ \bibinfo {pages} {064901}
  (\bibinfo {year} {2005})},\ \Eprint {http://arxiv.org/abs/nucl-th/0411110}
  {arXiv:nucl-th/0411110 [nucl-th]} \BibitemShut {NoStop}%
\bibitem [{\citenamefont {Xu}\ and\ \citenamefont {Greiner}(2007)}]{Xu:2007aa}%
  \BibitemOpen
  \bibfield  {author} {\bibinfo {author} {\bibfnamefont {Z.}~\bibnamefont
  {Xu}}\ and\ \bibinfo {author} {\bibfnamefont {C.}~\bibnamefont {Greiner}},\
  }\href {\doibase 10.1103/PhysRevC.76.024911} {\bibfield  {journal} {\bibinfo
  {journal} {Phys.Rev.}\ }\textbf {\bibinfo {volume} {C76}},\ \bibinfo {pages}
  {024911} (\bibinfo {year} {2007})},\ \Eprint
  {http://arxiv.org/abs/hep-ph/0703233} {arXiv:hep-ph/0703233} \BibitemShut
  {NoStop}%
\bibitem [{\citenamefont {Uphoff}\ \emph {et~al.}(2011)\citenamefont {Uphoff},
  \citenamefont {Fochler}, \citenamefont {Xu},\ and\ \citenamefont
  {Greiner}}]{Uphoff:2011ad}%
  \BibitemOpen
  \bibfield  {author} {\bibinfo {author} {\bibfnamefont {J.}~\bibnamefont
  {Uphoff}}, \bibinfo {author} {\bibfnamefont {O.}~\bibnamefont {Fochler}},
  \bibinfo {author} {\bibfnamefont {Z.}~\bibnamefont {Xu}}, \ and\ \bibinfo
  {author} {\bibfnamefont {C.}~\bibnamefont {Greiner}},\ }\href {\doibase
  10.1103/PhysRevC.84.024908} {\bibfield  {journal} {\bibinfo  {journal}
  {Phys.Rev.}\ }\textbf {\bibinfo {volume} {C84}},\ \bibinfo {pages} {024908}
  (\bibinfo {year} {2011})},\ \Eprint {http://arxiv.org/abs/1104.2295}
  {arXiv:1104.2295 [hep-ph]} \BibitemShut {NoStop}%
\bibitem [{\citenamefont {Uphoff}\ \emph {et~al.}(2012)\citenamefont {Uphoff},
  \citenamefont {Fochler}, \citenamefont {Xu},\ and\ \citenamefont
  {Greiner}}]{Uphoff:2012gb}%
  \BibitemOpen
  \bibfield  {author} {\bibinfo {author} {\bibfnamefont {J.}~\bibnamefont
  {Uphoff}}, \bibinfo {author} {\bibfnamefont {O.}~\bibnamefont {Fochler}},
  \bibinfo {author} {\bibfnamefont {Z.}~\bibnamefont {Xu}}, \ and\ \bibinfo
  {author} {\bibfnamefont {C.}~\bibnamefont {Greiner}},\ }\href {\doibase
  10.1016/j.physletb.2012.09.069} {\bibfield  {journal} {\bibinfo  {journal}
  {Phys.Lett.}\ }\textbf {\bibinfo {volume} {B717}},\ \bibinfo {pages} {430}
  (\bibinfo {year} {2012})},\ \Eprint {http://arxiv.org/abs/1205.4945}
  {arXiv:1205.4945 [hep-ph]} \BibitemShut {NoStop}%
\bibitem [{\citenamefont {Sjostrand}\ \emph {et~al.}(2006)\citenamefont
  {Sjostrand}, \citenamefont {Mrenna},\ and\ \citenamefont
  {Skands}}]{Sjostrand:2006za}%
  \BibitemOpen
  \bibfield  {author} {\bibinfo {author} {\bibfnamefont {T.}~\bibnamefont
  {Sjostrand}}, \bibinfo {author} {\bibfnamefont {S.}~\bibnamefont {Mrenna}}, \
  and\ \bibinfo {author} {\bibfnamefont {P.}~\bibnamefont {Skands}},\
  }\href@noop {} {\bibfield  {journal} {\bibinfo  {journal} {JHEP}\ }\textbf
  {\bibinfo {volume} {05}},\ \bibinfo {pages} {026} (\bibinfo {year} {2006})},\
  \Eprint {http://arxiv.org/abs/hep-ph/0603175} {arXiv:hep-ph/0603175}
  \BibitemShut {NoStop}%
\bibitem [{\citenamefont {Uphoff}\ \emph {et~al.}(2010)\citenamefont {Uphoff},
  \citenamefont {Fochler}, \citenamefont {Xu},\ and\ \citenamefont
  {Greiner}}]{Uphoff:2010sh}%
  \BibitemOpen
  \bibfield  {author} {\bibinfo {author} {\bibfnamefont {J.}~\bibnamefont
  {Uphoff}}, \bibinfo {author} {\bibfnamefont {O.}~\bibnamefont {Fochler}},
  \bibinfo {author} {\bibfnamefont {Z.}~\bibnamefont {Xu}}, \ and\ \bibinfo
  {author} {\bibfnamefont {C.}~\bibnamefont {Greiner}},\ }\href {\doibase
  10.1103/PhysRevC.82.044906} {\bibfield  {journal} {\bibinfo  {journal}
  {Phys.Rev.}\ }\textbf {\bibinfo {volume} {C82}},\ \bibinfo {pages} {044906}
  (\bibinfo {year} {2010})},\ \Eprint {http://arxiv.org/abs/1003.4200}
  {arXiv:1003.4200 [hep-ph]} \BibitemShut {NoStop}%
\bibitem [{\citenamefont {Fochler}\ \emph {et~al.}(2013)\citenamefont
  {Fochler}, \citenamefont {Uphoff}, \citenamefont {Xu},\ and\ \citenamefont
  {Greiner}}]{Fochler:2013epa}%
  \BibitemOpen
  \bibfield  {author} {\bibinfo {author} {\bibfnamefont {O.}~\bibnamefont
  {Fochler}}, \bibinfo {author} {\bibfnamefont {J.}~\bibnamefont {Uphoff}},
  \bibinfo {author} {\bibfnamefont {Z.}~\bibnamefont {Xu}}, \ and\ \bibinfo
  {author} {\bibfnamefont {C.}~\bibnamefont {Greiner}},\ }\href {\doibase
  10.1103/PhysRevD.88.014018} {\bibfield  {journal} {\bibinfo  {journal}
  {Phys.Rev.}\ }\textbf {\bibinfo {volume} {D88}},\ \bibinfo {pages} {014018}
  (\bibinfo {year} {2013})},\ \Eprint {http://arxiv.org/abs/1302.5250}
  {arXiv:1302.5250 [hep-ph]} \BibitemShut {NoStop}%
\bibitem [{\citenamefont {Gunion}\ and\ \citenamefont
  {Bertsch}(1982)}]{Gunion:1981qs}%
  \BibitemOpen
  \bibfield  {author} {\bibinfo {author} {\bibfnamefont {J.}~\bibnamefont
  {Gunion}}\ and\ \bibinfo {author} {\bibfnamefont {G.}~\bibnamefont
  {Bertsch}},\ }\href {\doibase 10.1103/PhysRevD.25.746} {\bibfield  {journal}
  {\bibinfo  {journal} {Phys.Rev.}\ }\textbf {\bibinfo {volume} {D25}},\
  \bibinfo {pages} {746} (\bibinfo {year} {1982})}\BibitemShut {NoStop}%
\bibitem [{\citenamefont {Xu}\ \emph {et~al.}(2008)\citenamefont {Xu},
  \citenamefont {Greiner},\ and\ \citenamefont {St\"ocker}}]{Xu:2007jv}%
  \BibitemOpen
  \bibfield  {author} {\bibinfo {author} {\bibfnamefont {Z.}~\bibnamefont
  {Xu}}, \bibinfo {author} {\bibfnamefont {C.}~\bibnamefont {Greiner}}, \ and\
  \bibinfo {author} {\bibfnamefont {H.}~\bibnamefont {St\"ocker}},\ }\href
  {\doibase 10.1103/PhysRevLett.101.082302} {\bibfield  {journal} {\bibinfo
  {journal} {Phys.Rev.Lett.}\ }\textbf {\bibinfo {volume} {101}},\ \bibinfo
  {pages} {082302} (\bibinfo {year} {2008})},\ \Eprint
  {http://arxiv.org/abs/0711.0961} {arXiv:0711.0961 [nucl-th]} \BibitemShut
  {NoStop}%
\bibitem [{\citenamefont {Xu}\ and\ \citenamefont {Greiner}(2009)}]{Xu:2008av}%
  \BibitemOpen
  \bibfield  {author} {\bibinfo {author} {\bibfnamefont {Z.}~\bibnamefont
  {Xu}}\ and\ \bibinfo {author} {\bibfnamefont {C.}~\bibnamefont {Greiner}},\
  }\href {\doibase 10.1103/PhysRevC.79.014904} {\bibfield  {journal} {\bibinfo
  {journal} {Phys.Rev.}\ }\textbf {\bibinfo {volume} {C79}},\ \bibinfo {pages}
  {014904} (\bibinfo {year} {2009})},\ \Eprint {http://arxiv.org/abs/0811.2940}
  {arXiv:0811.2940 [hep-ph]} \BibitemShut {NoStop}%
\bibitem [{\citenamefont {Xu}\ and\ \citenamefont {Greiner}(2010)}]{Xu:2010cq}%
  \BibitemOpen
  \bibfield  {author} {\bibinfo {author} {\bibfnamefont {Z.}~\bibnamefont
  {Xu}}\ and\ \bibinfo {author} {\bibfnamefont {C.}~\bibnamefont {Greiner}},\
  }\href {\doibase 10.1103/PhysRevC.81.054901} {\bibfield  {journal} {\bibinfo
  {journal} {Phys.Rev.}\ }\textbf {\bibinfo {volume} {C81}},\ \bibinfo {pages}
  {054901} (\bibinfo {year} {2010})},\ \Eprint {http://arxiv.org/abs/1001.2912}
  {arXiv:1001.2912 [hep-ph]} \BibitemShut {NoStop}%
\bibitem [{\citenamefont {Fochler}\ \emph {et~al.}(2011)\citenamefont
  {Fochler}, \citenamefont {Uphoff}, \citenamefont {Xu},\ and\ \citenamefont
  {Greiner}}]{Fochler:2011en}%
  \BibitemOpen
  \bibfield  {author} {\bibinfo {author} {\bibfnamefont {O.}~\bibnamefont
  {Fochler}}, \bibinfo {author} {\bibfnamefont {J.}~\bibnamefont {Uphoff}},
  \bibinfo {author} {\bibfnamefont {Z.}~\bibnamefont {Xu}}, \ and\ \bibinfo
  {author} {\bibfnamefont {C.}~\bibnamefont {Greiner}},\ }\href {\doibase
  10.1088/0954-3899/38/12/124152} {\bibfield  {journal} {\bibinfo  {journal}
  {J.Phys.}\ }\textbf {\bibinfo {volume} {G38}},\ \bibinfo {pages} {124152}
  (\bibinfo {year} {2011})},\ \Eprint {http://arxiv.org/abs/1107.0130}
  {arXiv:1107.0130 [hep-ph]} \BibitemShut {NoStop}%
\bibitem [{\citenamefont {Fochler}\ \emph {et~al.}(2009)\citenamefont
  {Fochler}, \citenamefont {Xu},\ and\ \citenamefont
  {Greiner}}]{Fochler:2008ts}%
  \BibitemOpen
  \bibfield  {author} {\bibinfo {author} {\bibfnamefont {O.}~\bibnamefont
  {Fochler}}, \bibinfo {author} {\bibfnamefont {Z.}~\bibnamefont {Xu}}, \ and\
  \bibinfo {author} {\bibfnamefont {C.}~\bibnamefont {Greiner}},\ }\href
  {\doibase 10.1103/PhysRevLett.102.202301} {\bibfield  {journal} {\bibinfo
  {journal} {Phys.Rev.Lett.}\ }\textbf {\bibinfo {volume} {102}},\ \bibinfo
  {pages} {202301} (\bibinfo {year} {2009})},\ \Eprint
  {http://arxiv.org/abs/0806.1169} {arXiv:0806.1169 [hep-ph]} \BibitemShut
  {NoStop}%
\bibitem [{\citenamefont {Fochler}\ \emph {et~al.}(2010)\citenamefont
  {Fochler}, \citenamefont {Xu},\ and\ \citenamefont
  {Greiner}}]{Fochler:2010wn}%
  \BibitemOpen
  \bibfield  {author} {\bibinfo {author} {\bibfnamefont {O.}~\bibnamefont
  {Fochler}}, \bibinfo {author} {\bibfnamefont {Z.}~\bibnamefont {Xu}}, \ and\
  \bibinfo {author} {\bibfnamefont {C.}~\bibnamefont {Greiner}},\ }\href
  {\doibase 10.1103/PhysRevC.82.024907} {\bibfield  {journal} {\bibinfo
  {journal} {Phys.Rev.}\ }\textbf {\bibinfo {volume} {C82}},\ \bibinfo {pages}
  {024907} (\bibinfo {year} {2010})},\ \Eprint {http://arxiv.org/abs/1003.4380}
  {arXiv:1003.4380 [hep-ph]} \BibitemShut {NoStop}%
\bibitem [{\citenamefont {Adare}\ \emph {et~al.}(2008)\citenamefont {Adare}
  \emph {et~al.}}]{Adare:2008qa}%
  \BibitemOpen
  \bibfield  {author} {\bibinfo {author} {\bibfnamefont {A.}~\bibnamefont
  {Adare}} \emph {et~al.} (\bibinfo {collaboration} {PHENIX Collaboration}),\
  }\href {\doibase 10.1103/PhysRevLett.101.232301} {\bibfield  {journal}
  {\bibinfo  {journal} {Phys.Rev.Lett.}\ }\textbf {\bibinfo {volume} {101}},\
  \bibinfo {pages} {232301} (\bibinfo {year} {2008})},\ \Eprint
  {http://arxiv.org/abs/0801.4020} {arXiv:0801.4020 [nucl-ex]} \BibitemShut
  {NoStop}%
\bibitem [{\citenamefont {Albino}\ \emph {et~al.}(2008)\citenamefont {Albino},
  \citenamefont {Kniehl},\ and\ \citenamefont {Kramer}}]{Albino:2008fy}%
  \BibitemOpen
  \bibfield  {author} {\bibinfo {author} {\bibfnamefont {S.}~\bibnamefont
  {Albino}}, \bibinfo {author} {\bibfnamefont {B.}~\bibnamefont {Kniehl}}, \
  and\ \bibinfo {author} {\bibfnamefont {G.}~\bibnamefont {Kramer}},\ }\href
  {\doibase 10.1016/j.nuclphysb.2008.05.017} {\bibfield  {journal} {\bibinfo
  {journal} {Nucl.Phys.}\ }\textbf {\bibinfo {volume} {B803}},\ \bibinfo
  {pages} {42} (\bibinfo {year} {2008})},\ \Eprint
  {http://arxiv.org/abs/0803.2768} {arXiv:0803.2768 [hep-ph]} \BibitemShut
  {NoStop}%
\bibitem [{\citenamefont {Abelev}\ \emph {et~al.}(2013)\citenamefont {Abelev}
  \emph {et~al.}}]{Abelev:2012hxa}%
  \BibitemOpen
  \bibfield  {author} {\bibinfo {author} {\bibfnamefont {B.}~\bibnamefont
  {Abelev}} \emph {et~al.} (\bibinfo {collaboration} {ALICE Collaboration}),\
  }\href {\doibase 10.1016/j.physletb.2013.01.051} {\bibfield  {journal}
  {\bibinfo  {journal} {Phys.Lett.}\ }\textbf {\bibinfo {volume} {B720}},\
  \bibinfo {pages} {52} (\bibinfo {year} {2013})},\ \Eprint
  {http://arxiv.org/abs/1208.2711} {arXiv:1208.2711 [hep-ex]} \BibitemShut
  {NoStop}%
\bibitem [{\citenamefont {Adams}\ \emph {et~al.}(2005)\citenamefont {Adams}
  \emph {et~al.}}]{Adams:2004bi}%
  \BibitemOpen
  \bibfield  {author} {\bibinfo {author} {\bibfnamefont {J.}~\bibnamefont
  {Adams}} \emph {et~al.} (\bibinfo {collaboration} {STAR Collaboration}),\
  }\href {\doibase 10.1103/PhysRevC.72.014904} {\bibfield  {journal} {\bibinfo
  {journal} {Phys.Rev.}\ }\textbf {\bibinfo {volume} {C72}},\ \bibinfo {pages}
  {014904} (\bibinfo {year} {2005})},\ \Eprint
  {http://arxiv.org/abs/nucl-ex/0409033} {arXiv:nucl-ex/0409033 [nucl-ex]}
  \BibitemShut {NoStop}%
\bibitem [{\citenamefont {Back}\ \emph {et~al.}(2005)\citenamefont {Back} \emph
  {et~al.}}]{Back:2004mh}%
  \BibitemOpen
  \bibfield  {author} {\bibinfo {author} {\bibfnamefont {B.}~\bibnamefont
  {Back}} \emph {et~al.} (\bibinfo {collaboration} {PHOBOS Collaboration}),\
  }\href {\doibase 10.1103/PhysRevC.72.051901} {\bibfield  {journal} {\bibinfo
  {journal} {Phys.Rev.}\ }\textbf {\bibinfo {volume} {C72}},\ \bibinfo {pages}
  {051901} (\bibinfo {year} {2005})},\ \Eprint
  {http://arxiv.org/abs/nucl-ex/0407012} {arXiv:nucl-ex/0407012 [nucl-ex]}
  \BibitemShut {NoStop}%
\bibitem [{\citenamefont {Chatrchyan}\ \emph {et~al.}(2013)\citenamefont
  {Chatrchyan} \emph {et~al.}}]{Chatrchyan:2012ta}%
  \BibitemOpen
  \bibfield  {author} {\bibinfo {author} {\bibfnamefont {S.}~\bibnamefont
  {Chatrchyan}} \emph {et~al.} (\bibinfo {collaboration} {CMS Collaboration}),\
  }\href {\doibase 10.1103/PhysRevC.87.014902} {\bibfield  {journal} {\bibinfo
  {journal} {Phys.Rev.}\ }\textbf {\bibinfo {volume} {C87}},\ \bibinfo {pages}
  {014902} (\bibinfo {year} {2013})},\ \Eprint {http://arxiv.org/abs/1204.1409}
  {arXiv:1204.1409 [nucl-ex]} \BibitemShut {NoStop}%
\bibitem [{\citenamefont {Auvinen}\ and\ \citenamefont
  {Petersen}(2013)}]{Auvinen:2013sba}%
  \BibitemOpen
  \bibfield  {author} {\bibinfo {author} {\bibfnamefont {J.}~\bibnamefont
  {Auvinen}}\ and\ \bibinfo {author} {\bibfnamefont {H.}~\bibnamefont
  {Petersen}},\ }\href@noop {} {\  (\bibinfo {year} {2013})},\ \Eprint
  {http://arxiv.org/abs/1310.1764} {arXiv:1310.1764 [nucl-th]} \BibitemShut
  {NoStop}%
\bibitem [{\citenamefont {Wesp}\ \emph {et~al.}(2011)\citenamefont {Wesp},
  \citenamefont {El}, \citenamefont {Reining}, \citenamefont {Xu},
  \citenamefont {Bouras} \emph {et~al.}}]{Wesp:2011yy}%
  \BibitemOpen
  \bibfield  {author} {\bibinfo {author} {\bibfnamefont {C.}~\bibnamefont
  {Wesp}}, \bibinfo {author} {\bibfnamefont {A.}~\bibnamefont {El}}, \bibinfo
  {author} {\bibfnamefont {F.}~\bibnamefont {Reining}}, \bibinfo {author}
  {\bibfnamefont {Z.}~\bibnamefont {Xu}}, \bibinfo {author} {\bibfnamefont
  {I.}~\bibnamefont {Bouras}},  \emph {et~al.},\ }\href {\doibase
  10.1103/PhysRevC.84.054911} {\bibfield  {journal} {\bibinfo  {journal}
  {Phys.Rev.}\ }\textbf {\bibinfo {volume} {C84}},\ \bibinfo {pages} {054911}
  (\bibinfo {year} {2011})},\ \Eprint {http://arxiv.org/abs/1106.4306}
  {arXiv:1106.4306 [hep-ph]} \BibitemShut {NoStop}%
\bibitem [{\citenamefont {Gale}\ \emph
  {et~al.}(2013{\natexlab{b}})\citenamefont {Gale}, \citenamefont {Jeon},
  \citenamefont {Schenke}, \citenamefont {Tribedy},\ and\ \citenamefont
  {Venugopalan}}]{Gale:2012rq}%
  \BibitemOpen
  \bibfield  {author} {\bibinfo {author} {\bibfnamefont {C.}~\bibnamefont
  {Gale}}, \bibinfo {author} {\bibfnamefont {S.}~\bibnamefont {Jeon}}, \bibinfo
  {author} {\bibfnamefont {B.}~\bibnamefont {Schenke}}, \bibinfo {author}
  {\bibfnamefont {P.}~\bibnamefont {Tribedy}}, \ and\ \bibinfo {author}
  {\bibfnamefont {R.}~\bibnamefont {Venugopalan}},\ }\href {\doibase
  10.1103/PhysRevLett.110.012302} {\bibfield  {journal} {\bibinfo  {journal}
  {Phys.Rev.Lett.}\ }\textbf {\bibinfo {volume} {110}},\ \bibinfo {pages}
  {012302} (\bibinfo {year} {2013}{\natexlab{b}})},\ \Eprint
  {http://arxiv.org/abs/1209.6330} {arXiv:1209.6330 [nucl-th]} \BibitemShut
  {NoStop}%
\bibitem [{\citenamefont {Zapp}\ \emph {et~al.}(2009)\citenamefont {Zapp},
  \citenamefont {Stachel},\ and\ \citenamefont {Wiedemann}}]{Zapp:2008af}%
  \BibitemOpen
  \bibfield  {author} {\bibinfo {author} {\bibfnamefont {K.}~\bibnamefont
  {Zapp}}, \bibinfo {author} {\bibfnamefont {J.}~\bibnamefont {Stachel}}, \
  and\ \bibinfo {author} {\bibfnamefont {U.~A.}\ \bibnamefont {Wiedemann}},\
  }\href {\doibase 10.1103/PhysRevLett.103.152302} {\bibfield  {journal}
  {\bibinfo  {journal} {Phys.Rev.Lett.}\ }\textbf {\bibinfo {volume} {103}},\
  \bibinfo {pages} {152302} (\bibinfo {year} {2009})},\ \Eprint
  {http://arxiv.org/abs/0812.3888} {arXiv:0812.3888 [hep-ph]} \BibitemShut
  {NoStop}%
\bibitem [{\citenamefont {Coleman-Smith}\ and\ \citenamefont
  {M\"uller}(2012)}]{ColemanSmith:2012vr}%
  \BibitemOpen
  \bibfield  {author} {\bibinfo {author} {\bibfnamefont {C.~E.}\ \bibnamefont
  {Coleman-Smith}}\ and\ \bibinfo {author} {\bibfnamefont {B.}~\bibnamefont
  {M\"uller}},\ }\href {\doibase 10.1103/PhysRevC.86.054901} {\bibfield
  {journal} {\bibinfo  {journal} {Phys.Rev.}\ }\textbf {\bibinfo {volume}
  {C86}},\ \bibinfo {pages} {054901} (\bibinfo {year} {2012})},\ \Eprint
  {http://arxiv.org/abs/1205.6781} {arXiv:1205.6781 [hep-ph]} \BibitemShut
  {NoStop}%
\bibitem [{\citenamefont {Senzel}\ \emph {et~al.}(2013)\citenamefont {Senzel},
  \citenamefont {Fochler}, \citenamefont {Uphoff}, \citenamefont {Xu},\ and\
  \citenamefont {Greiner}}]{Senzel:2013dta}%
  \BibitemOpen
  \bibfield  {author} {\bibinfo {author} {\bibfnamefont {F.}~\bibnamefont
  {Senzel}}, \bibinfo {author} {\bibfnamefont {O.}~\bibnamefont {Fochler}},
  \bibinfo {author} {\bibfnamefont {J.}~\bibnamefont {Uphoff}}, \bibinfo
  {author} {\bibfnamefont {Z.}~\bibnamefont {Xu}}, \ and\ \bibinfo {author}
  {\bibfnamefont {C.}~\bibnamefont {Greiner}},\ }\href@noop {} {\  (\bibinfo
  {year} {2013})},\ \Eprint {http://arxiv.org/abs/1309.1657} {arXiv:1309.1657
  [hep-ph]} \BibitemShut {NoStop}%
\bibitem [{\citenamefont {Uphoff}\ \emph {et~al.}(2013)\citenamefont {Uphoff},
  \citenamefont {Senzel}, \citenamefont {Xu},\ and\ \citenamefont
  {Greiner}}]{Uphoff:2013rka}%
  \BibitemOpen
  \bibfield  {author} {\bibinfo {author} {\bibfnamefont {J.}~\bibnamefont
  {Uphoff}}, \bibinfo {author} {\bibfnamefont {F.}~\bibnamefont {Senzel}},
  \bibinfo {author} {\bibfnamefont {Z.}~\bibnamefont {Xu}}, \ and\ \bibinfo
  {author} {\bibfnamefont {C.}~\bibnamefont {Greiner}},\ }\href@noop {} {\
  (\bibinfo {year} {2013})},\ \Eprint {http://arxiv.org/abs/1310.1340}
  {arXiv:1310.1340 [hep-ph]} \BibitemShut {NoStop}%
\bibitem [{\citenamefont {Abir}\ \emph {et~al.}(2012)\citenamefont {Abir},
  \citenamefont {Greiner}, \citenamefont {Martinez}, \citenamefont {Mustafa},\
  and\ \citenamefont {Uphoff}}]{Abir:2011jb}%
  \BibitemOpen
  \bibfield  {author} {\bibinfo {author} {\bibfnamefont {R.}~\bibnamefont
  {Abir}}, \bibinfo {author} {\bibfnamefont {C.}~\bibnamefont {Greiner}},
  \bibinfo {author} {\bibfnamefont {M.}~\bibnamefont {Martinez}}, \bibinfo
  {author} {\bibfnamefont {M.~G.}\ \bibnamefont {Mustafa}}, \ and\ \bibinfo
  {author} {\bibfnamefont {J.}~\bibnamefont {Uphoff}},\ }\href {\doibase
  10.1103/PhysRevD.85.054012} {\bibfield  {journal} {\bibinfo  {journal}
  {Phys.Rev.}\ }\textbf {\bibinfo {volume} {D85}},\ \bibinfo {pages} {054012}
  (\bibinfo {year} {2012})},\ \Eprint {http://arxiv.org/abs/1109.5539}
  {arXiv:1109.5539 [hep-ph]} \BibitemShut {NoStop}%
\end{thebibliography}%

\end{document}